\documentclass[11pt]{article}
\usepackage{graphicx,a4,epsfig}
\newcommand{\be}{\begin{equation}}
\newcommand{\en}{\end{equation}}
\newcommand{\bea}{\begin{eqnarray}}
\newcommand{\ena}{\end{eqnarray}}
\newcommand{\lbl}[1]{\label{eq:#1}}
\newcommand{\rf}[1]{(\ref{eq:#1})}

\def\today{\ifcase \month\or
  January\or February\or March\or April\or May\or June\or
    July\or August\or September\or October\or November\or December\fi
      \space\number\day,\space
        \number\year }
\newcommand{\lapprox}{%
\mathrel{%
\setbox0=\hbox{$<$}\raise0.6ex\copy0\kern-\wd0\lower0.65ex\hbox{$\sim$}}}

\newcommand{\gapprox}{%
\mathrel{%
\setbox0=\hbox{$>$}\raise0.6ex\copy0\kern-\wd0\lower0.65ex\hbox{$\sim$}}}

\def\mpd{{m^2_+}}
\def\mmd{{m^2_-}}
\def\cut{{\Lambda^2}}

\def\Del{\Delta_{K \pi}}
\def\Dkpi{\Delta_{K \pi}}
\def\Dketa{\Delta_{K \eta}}
\def\Skpi{\Sigma_{\pi K}}
\def\Sketa{\Sigma_{K \eta}}
\def\Jpik{\bar J_{\pi K}(s)}
\def\Jketa{\bar J_{K \eta}(s)}
\def\JJketa{\bar J'_{K \eta}(0)}
\def\JJpik{\bar J'_{\pi K}(0)}
\def\Jpipi{\bar J_{\pi\pi}(t)}
\def\Jkk{\bar J_{KK}(t)}
\def\upi{{1\over\pi}}
\begin{document}
\vskip -1 truecm
\rightline{IISc-CTS-12/01}
\rightline{FZJ-IKP(Th)-2001-10}
\rightline{IPNO-DR 01-014}
\rightline{\today}
\begin{center}
{\bf $\pi K$ SUM RULES AND THE SU(3) CHIRAL EXPANSION}\\
\bigskip
{\bf B. Ananthanarayan}\\
{\sl Centre for Theoretical Studies}\\
{\sl Indian Institute of Science}\\
{\sl Bangalore 560 012, India}

{\bf P. B\"uttiker}\\
{\sl Institut f\"ur Kernphysik}\\
{\sl Forschungszentrum J\"ulich}\\
{\sl D-52425 J\"ulich, Germany}

{\bf B. Moussallam}\\
{\sl Groupe de Physique Th\'eorique, IPN }\\
{\sl Universit\'e Paris-Sud}\\
{\sl F-91406 Orsay C\'edex, France}
\end{center}
\medskip
\vskip 0.8truecm
\centerline{\large\bf Abstract}

A recently proposed set of sum rules, based on the pion-Kaon 
scattering amplitudes and their crossing-symmetric conjugates are
analysed in detail. A key role is played by the $l=0$
$\pi\pi\to K\overline K$ amplitude which requires an extrapolation to be 
performed. It is shown how this is tightly constrained from analyticity,
chiral counting and the available experimental data, and  its stability
is tested. A re-evaluation of the $O(p^4)$ chiral couplings
$L_1$, $L_2$, $L_3$ is obtained, as well as a new 
evaluation of the large $N_c$ suppressed coupling $L_4$.

\newpage
\noindent{\large\bf 1. Introduction}

Chiral perturbation theory is a rigorous approach to QCD in a 
restricted but nonperturbative regime, which has recently been
developed to $O(p^6)$, i.e. to the next-to-next to the leading 
order \cite{op6}. Foundations of this method \cite{W} and the abundant work
which has followed the basic papers where the NLO  theory was
defined \cite{gl84,gl85} are summarized in the 
review \cite{revs}. The chiral expansion is based on an 
effective field theory
and, as such,  involves an increasing number of coupling constants
with increasing chiral order. In $SU(3)$, ten couplings $L_i(\mu)$
are involved at $O(p^4)$ and ninety more couplings $C_i(\mu)$ appear 
at the next order. In order to make predictions to $O(p^6)$ accuracy, 
estimates of the $C_i(\mu)$ must be performed but, moreover, the values
of the $L_i$ may have to be modified, compared to their determination
using $O(p^4)$ accuracy. The order of magnitude of 
such a variation, which reflects the rate of convergence of the expansion
in the strange quark mass,  can be estimated by comparing several different
$O(p^4)$ determinations  of the same coupling constants. This is one purpose 
of the present work in which we propose a new determination of
$L_1$, $L_2$ and $L_3$ from a set of sum rules based on the pion-Kaon
amplitude and its expression in ChPT at one-loop \cite{bkm}. 
We will compare these results
with the previous determination from the $K_{l4}$ 
form-factors \cite{Kl4bij,Kl4rigg} and the (partial) 
determination from $\pi\pi$ sum rules \cite{gl84}. 

A few $O(p^4)$ coupling constants are still very poorly known, 
in particular, those which are suppressed in the large $N_c$ limit: $L_4$
and $L_6$. Naively, one may even question whether such a suppression 
should actually hold, because these couplings were shown to
be controlled by physics of the scalar meson resonances \cite{egpr}
which fail to obey simple large $N_c$ rules. On a more 
sophisticated level, one may
note that some of  the  large $N_c$ suppressed mechanisms, like 
internal quark loops, are partly taken into account in the chiral expansion
via meson loops. The question remains of what value of 
the scale $\mu$ is the one
at which the suppression operates. Another related interesting issue
is that of the phase structure of QCD-like theories as a function 
of the number $N_F$ of massless flavours and the value of
$N^{crit}_F$ above which chiral symmetry is no longer spontaneously
broken. Some recent lattice simulations \cite{latt} 
have obtained values as small as  $N^{crit}_F\simeq 4$. If true, this should
affect the $SU(3)$ chiral expansion.
For instance, it can be seen that $L_4$
and $L_6$ control how the chiral order parameters 
$F_\pi$ and $<\bar u u>$ respectively 
evolve from $N_F=2$ to $N_F=3$ \cite{gl85}.
Clearly, a small value of $N^{crit}_F$,  should lead to
anomalously large values of $L_4,\  L_6$. 

In view of this, an interesting outcome of the present work is a determination
of $L_4$. In principle,  
it could have been extracted  from the $K_{l4}$ form-factors, but this is not 
feasible in practice because its contribution  is accidentally
suppressed \cite{Kl4bij,Kl4rigg}. 
Here, we will take advantage of the fact that no such 
suppression affects the $\pi K$ amplitude and we will show that 
an evaluation of $L_4$ is then possible, which is at the same level of 
reliability and accuracy as that of $L_1,\ L_2,\ L_3$. Several recent
papers have considered aspects of the pion-Kaon scattering 
amplitudes \cite{black,roessl,jop}. One purpose is a better
understanding of the scalar resonances (see e.g. \cite{mink} for a
recent discussion of the experimental situation). 
This question, of course, is
not unrelated to that of the size of the chiral couplings \cite{egpr}.

The dispersive formalism on which the sum rules are based has been 
developed in a previous paper \cite{ab}. This 
formalism is reviewed in sec.2 below and
presented in a form suitable for comparison with the $O(p^4)$ expression
of the amplitude, which has been computed some time ago 
by Bernard et al. \cite{bkm},
as well as the $O(p^6)$ expression which should be available in
the near future. The detailed form of the sum rules for the $O(p^4)$ 
coupling constants are then presented in sec.3. 
The practical evaluation of these sum rules, making use of the available
experimental data then presents a difficulty: because of $s-t$ crossing the
$\pi\pi\to K\overline K$ amplitude appears and it is needed below 
the experimentally accessible energy range. This was noted in ref. \cite{ab}
in which only results not depending on this amplitude were presented. 
Extrapolation of the
$\pi\pi\to K\overline K$ amplitude, in particular for the S-wave, 
is a problem which was considered a long time ago \cite{J&P,J&N}. We
discuss this question in some detail in sec.4, 
and then present all the results.

\noindent{\large\bf 2. Dispersive representation, crossing-symmetry 
and chiral counting}

Basic work on dispersion relations related to the pion-Kaon amplitudes
has been reviewed by Lang \cite{lang}. In order to determine the number of
subtractions we make the assumption that standard Regge phenomenology
applies.
As shown in ref. \cite{ab} the dispersive
representation can be recast in a specific form by taking into account chiral
counting. Dropping terms which are of chiral order $O(p^8)$ it can be put in
a form which involves functions of only one of the Mandelstam 
variables $s,\ t,\ u$ and are analytic, except for a right-hand cut, plus
a polynomial. This was first demonstrated for the case of the pion-pion
amplitude in ref. \cite{ssf}. Let us begin by recalling some basic facts and
some notation. 

\noindent{\bf 2.1 Notation and conventions}

Making use of $s-u$ crossing, the two independent isospin $I=1/2$
and $I=3/2$ pion-Kaon amplitudes can be expressed in terms of the $I=3/2$ one,
\be
F^{1\over2}(s,t,u)=-{1\over2}F^{3\over2}(s,t,u)+
                    {3\over2}F^{3\over2}(u,t,s)\ .
\en
It is convenient to introduce the amplitudes $F^+$ and 
$F^-$ which are respectively even and odd under $s-u$ crossing 
because they require a different number of subtractions. In terms of isospin
amplitudes, they are defined as
\bea
&&F^+(s,t,u)={1\over3} F^{1\over2}(s,t,u)
+            {2\over3} F^{3\over2}(s,t,u)\nonumber\\
&&F^-(s,t,u)={1\over3} F^{1\over2}(s,t,u)
-            {1\over3} F^{3\over2}(s,t,u)\ .
\ena
Under $s-t$ crossing, one generates the $I=0$ and $I=1$ 
$\pi\pi\to K\overline K$
amplitudes,
\bea
&&G^0(t,s,u)=\sqrt6 F^+(s,t,u)\nonumber\\
&&G^1(t,s,u)=2      F^-(s,t,u)\ .
\ena
The partial wave expansion of the $\pi K$ isospin amplitudes are defined
as
\be\lbl{Fpw}
F^I(s,t)=16\pi\sum_l (2 l+1) P_l(z_s) f^I_l(s)\ .
\en
In a similar way we can expand $F^+$ and $F^-$, the corresponding 
partial-wave projections are denoted $f^+_l(s)$ and $f^-_l(s)$. The s-channel
scattering angle appearing above is given by
\be
z_s={s(t-u)+m^2_-m^2_+\over(s-m^2_-)(s-m^2_+)}
\quad {\rm with}\ m_\pm=m_K\pm m_\pi\ .
\en
The partial-wave expansion of the $\pi\pi\to K\overline K$ amplitude 
is conventionally defined as,
\be\lbl{Gpw}
G^I(t,s)=16\pi\sqrt2\,\sum_l (2 l+1) P_l(z_t) g^{I}_l(t) (q_\pi(t) 
q_K(t))^l\ .
\en
with
\be
q_P(t)={1\over 2}\sqrt{t-4m^2_P},\quad
z_t={s-u\over 4 q_\pi(t)q_K(t)}\ .
\en
The rationale for introducing the factor $(q_\pi q_K)^l$ in eq. \rf{Gpw} is
explained by Frazer and Fulco \cite{ff}. It ensures that the partial-wave 
amplitudes $g^{I}_l(t)$ have good analytic properties. 
With these definitions, the partial-wave S-matrices are given by
\bea
&&\pi K\to\pi K:  S^I_l(s)=1+2i{\sqrt{(s-\mmd)(s-\mpd)}\over s} f^I_l(s)
\nonumber\\
&&\pi\pi\to K\overline K: S^I_l(t)=4i
{(q_\pi(t) q_K(t))^{l+1/2}\over\sqrt{t}} g^I_l(t)\ .
\ena

\noindent{\bf 2.2 Dispersive representation of $F^+(s,t)$}

One first writes down a dispersive representation with $t$ fixed (and small).
According to Regge phenomenology, the asymptotic dependence as a function of 
$s$ is controlled by the Pomeron, implying the need for two subtractions, which
would also result on the general basis of the Froissart bound,
\be\lbl{dr}
F^+(s,t)=\tilde c(t)+{1\over\pi}\int^\infty_\mpd {ds'\over (s')^2}\left(
{s^2\over s'-s} +{u^2\over s'-u}\right) Im F^+(s',t) ,
\en
giving $F^+(s,t)$ in terms of an unknown function of $t$. Next, 
following ref. \cite{ssf}, one splits
the integration range into two regions a) $[\mpd,\Lambda^2]$ and 
b)$[\Lambda^2, \infty]$, $\Lambda$ being the scale of the chiral expansion,
i.e. $\Lambda\simeq 1$ GeV. In the lower integration range, we can apply
the chiral counting and drop the imaginary parts of the partial waves
with $l\ge 2$ which are  $O(p^8)$, i.e. we put 
\be
Im F^+(s',t)=16\pi \left[
Im f_0^+ +3 Im f_1^+(s'){s'(t-u')+\Del^2\over(s'-\mmd)(s'-\mpd)}\right],\
s'<\Lambda^2\ .
\en
In the region b) we can expand in terms of $s,\ t,\ u$ divided by
$\Lambda^2$ again dropping terms which are $O(p^8)$. After some reshuffling
of part a) and absorbing functions of $t$ into $\tilde c(t)$, one obtains
the fixed $t$ dispersive representation in the form
\bea\lbl{tfix}
&&F^+(s,t)= c(t) +\left[ W_0^+(s)+(t-u)W_1^+(s) + (s\leftrightarrow u)\right]
\nonumber\\
&& -(2us){1\over\pi}\int_{\Lambda^2}^\infty 
{ds'\over (s')^3}\left( 1+3{\Skpi-t/2\over s'}\right)Im F^+(s',t)
+O(p^8)
\ena
In the equations above we have introduced the notation
\be
\Sigma_{PQ}=m^2_P+ m^2_Q,\quad
\Delta_{PQ}=m^2_P- m^2_Q\ .
\en
The functions $W_0^+(s)$, $W_1^+(s)$ are analytic except
for a right-hand cut and are given in terms of the $S$ and $P$ waves
of the pion-Kaon amplitude,
\bea\lbl{wplusdef}
&&W_0^+(s)=16\int_\mpd^{\Lambda^2} {ds'\over s'-s}
\left(  Im f_0^+(s')+
\Del^2{3 Im f_1^+(s')\over(s'-\mmd)(s'-\mpd)}\right)\nonumber\\
&&W_1^+(s)=16 s \int_\mpd^{\Lambda^2} {ds'\over s'-s} 
{3 Im f_1^+(s')\over(s'-\mmd)(s'-\mpd)}\ .
\ena

In order to further constrain the function $c(t)$ appearing in eq. \rf{tfix}
we must write down for $F^+(s,t)$ a dispersion relation 
involving the cut in the $t$ variable. A possibility is to use
a dispersion relation with $s$ fixed. Alternatively, one can
use one with $us$ fixed, $us=b$, (hyperbolic dispersion relation) which
was shown to have better convergence properties \cite{hyper}. 
In this case, the variables
$s$ and $u$ are functions of $t$ denoted $s_b$ and $u_b$,
\be\lbl{sb}
s_b(t)=\Skpi-{t\over2}+\sqrt{ \left(\Skpi-{t\over2}\right)^2-b},\ 
u_b(t)=b/s_b(t)\ .
\en
The function $F^+(s_b,t)$ is an analytic function of $t$ with
a) a right-hand cut $4m^2_\pi\le t < \infty$, and 
b) a left-hand cut $-\infty < t \le \mmd-b/\mpd$ .
In the following, we will adopt a specific value for $b$,
\be\lbl{bval} 
b\equiv\Dkpi^2=\mmd\mpd,
\en
which corresponds to backward scattering, $z_s=-1$. In that case, the upper 
limit of the left-hand cut is $t=0$. 
In the asymptotic regions $t\to\pm\infty$, it is simple to verify that 
the dominant divergence is controlled by the  
$K^*$ or $K^*_2$ Regge trajectories, and it is  therefore plausible 
that a {\sl single} subtraction is sufficient in this case. 
The following representation is then obtained \cite{J&N,ab},
\bea\lbl{usfix} 
F^+(s_b,t)=&&c_b+{1\over\pi}\int_{\mpd}^\infty {ds'\over s'}
\left( {s_b\over s'-s_b}+{u_b\over s'-u_b}\right)Im F^+(s',t'_b)
\nonumber\\ 
&&+{t\over\sqrt6\pi}\int_{4m^2_\pi}^\infty {dt' Im G^0(t',s'_b)
\over t'(t'-t)}\ ,
\ena
where $s'_b\equiv s_b(t')$ (see eq. \rf{sb} )and
\be
t'_b=2\Skpi-s'-{b\over s'}\ .
\en
Next, one splits the integration range as before and drops terms which
are $O(p^8)$. Equating the representations  \rf{tfix} and  \rf{usfix}
then determines the unknown function in the former expression 
leaving just one undetermined constant. 
Introducing the following notation for the
various high-energy integrals which are involved,
\bea\lbl{heplus}
&& H^+(n)     =\upi\int_{\Lambda^2}^\infty {ds'\over (s')^n }
Im F^+(s',0)\nonumber\\
&& \dot H^+(n)=\upi\int_{\Lambda^2}^\infty {ds'\over (s')^n }
\partial_t Im F^+(s',0)\nonumber\\
&& H^+_b(n)     =\upi\int_{\Lambda^2}^\infty {ds'\over (s')^n }
Im F^+(s',t'_b)\nonumber\\
&& G^+_b(n)     ={1\over\sqrt6\pi}
                  \int_{\Lambda^2}^\infty {dt'\over (t')^n }
Im G^0(t',s'_b)\ ,
\ena
we finally obtain the following dispersive representation for the amplitude
$F^+(s,t)$ :
\bea\lbl{drplus}
&&F^+(s,t)=C+\left[ W^+_0(s)+(t-u)W^+_1(s) +(s\leftrightarrow u)\right]+U_0(t)
\nonumber\\
&& +16t\int_\mpd^\cut ds' {3Im f_1^+(s')\over(s'-\mmd)(s'-\mpd)}
-2us\left[ H^+(3)+t\dot H^+(3)+3(\Skpi-t/2) H^+(4)\right]
\nonumber\\
&&+2bt\left[\dot H^+(3)-3/2  H^+(4)\right]
+t G^+_b(2) +t^2  G^+_b(3) +t^3  G^+_b(4)
\nonumber\\
&&-t H^+_b(2)+t(t-4\Skpi)H^+_b(3)-t(t^2-6t\Skpi+12\Skpi^2-3b)H^+_b(4)
+O(p^8)\ .
\ena
Apart from a polynomial, 
this expression involves the functions $W^+_0(z)$, 
$W^+_1(z)$ which are defined 
in terms of the $S$ and $P$ waves of the $\pi K$ amplitude 
(see eqs. \rf{wplusdef}\rf{wmindef}) 
and the function $U_0(z)$ which is defined in terms
of the $S$ wave of the $\pi\pi\to K\overline K$ amplitude,
\be
U_0(z)={16\over\sqrt3}\,z \int_{4m^2_\pi}^{\cut} dt'\,{Im g_0^0(t')
\over t'(t'-z)}\ .
\en
This derivation shows that the specific form of the amplitude
eq. \rf{drplus} must hold in chiral perturbation theory at $O(p^4)$
(which we will check explicitly below) and also at $O(p^6)$.

\noindent{\bf 2.3 Dispersive representation of $F^-(s,t)$}

We proceed in the same way as for $F^+(s,t)$ by first writing down
a dispersion relation with $t$ fixed, the only difference is that
now, the behaviour at large $s$ is dominated by the $K^*$ and $K^*_2$ Regge
exchanges and, therefore, a dispersion representation with no subtraction
should converge,
\be
F^-(s,t)={1\over\pi}\int^\infty_\mpd {ds'}\left(
{1\over s'-s} -{1\over s'-u}\right) Im F^-(s',t) .
\en
As before, one splits the integration range into two pieces and in
the lower energy range one retains only the $S$ and $P$ waves, obtaining
\bea\lbl{drmtfix}
&&F^-(s,t)=W_0^-(s)-W_0^-(u)+(t-u)W_1^-(s)-(t-s)W_1^-(u)
\nonumber\\
&&+(s-u)\left[
16\int_\mpd^\cut {3Im f_1^-(s')ds'\over(s'-\mmd)(s'-\mpd)}
+\upi\int_\cut^\infty {Im F^-(s',t)ds'\over(s'-s)(s'-u)}\right]\ .
\ena 
The functions $W_0^-(s)$ and $W_1^-(s)$ are exactly analogous to their $+$
counterparts defined above,
\bea\lbl{wmindef}
&&W_0^-(s)=16\int_\mpd^{\Lambda^2} {ds'\over s'-s}
\left(  Im f_0^-(s')+
\Del^2{3 Im f_1^-(s')\over(s'-\mmd)(s'-\mpd)}\right)\nonumber\\
&&W_1^-(s)=16\,s \int_\mpd^{\Lambda^2} {ds'\over s'-s} 
{3 Im f_1^-(s')\over(s'-\mmd)(s'-\mpd)}\ .
\ena
This representation has no undefined functions but convergence
is ensured only for negative values of $t$.
One can extend the range of validity in $t$, and also display
the cut structure by combining with a hyperbolic dispersion
relation. One writes a  dispersion
relation  at fixed $us=b$ for the function 
\be
F^-(s,t)/(s-u)
\en
which is even in $s-u$ and thus free of kinematical singularities,
and one obtains
\be\lbl{fminhyp}
{F^-(s_b,t)\over s_b-u_b}={1\over2\pi}\int_{4m^2_\pi}^\infty
{dt'\over t'-t}Im\left[ {G^1(t',s'_b)\over s'_b-u'_b}\right]
+{1\over\pi}\int_\mpd^\infty ds'{Im F^-(s',t'_b)\over(s'-s_b)(s'-u_b)}\ .
\en 
In the low energy region of the right-hand cut, only the P wave
of the $\pi\pi\to K \overline K$ amplitude will contribute, which
generates the function
\be
U_1(z)=6\sqrt2 \int_{4m_\pi^2}^\cut dt'
{Im g_1^1(t')\over t'-z}\ .
\en
Equating the fixed $t$ and fixed $us$ representations gives the
following equation, valid for $t\le0$,
\bea\lbl{relp}
&&
32\int_\mpd^\cut ds'{3 Im f_1^-(s')\over(s'-\mmd)(s'-\mpd)}
+\upi\int_\cut^\infty ds'{Im F^-(s',t)\over(s'-s_b)(s'-u_b)}=
\nonumber\\
&&U_1(t)
+{1\over2\pi}\int_\cut^\infty dt'{Im G^1(t',s'_b)\over(t'-t)(s'_b-u'_b)}
+\upi         \int_\cut^\infty ds'{Im F^-(s',t'_b)\over(s'-s_b)(s'-u_b)}
\ena
which relates the $P$-waves in the $\pi K$ and the $\pi\pi\to K\overline K$
channels. Finally, introducing the following notation
for the high-energy integrals,
\bea\lbl{hemin}
&& H^-(n)     =\upi\int_{\Lambda^2}^\infty {ds'\over (s')^n }
Im F^-(s',0)\nonumber\\
&& H^-_b(n)     =\upi\int_{\Lambda^2}^\infty {ds'\over (s')^n }
Im F^-(s',t'_b)\nonumber\\
&& G^-_b(n)     ={1\over2\pi}
                  \int_{\Lambda^2}^\infty {dt'\over (t')^{n-1}(s'_b-u'_b) }
Im G^1(t',s'_b)\ ,
\ena
we obtain the dispersive representation for $F^-(s,t)$, valid up to
$O(p^8)$ contributions, in the form,
\bea\lbl{drmin}
&&F^-(s,t)=W_0^-(s)-W_0^-(u)+(t-u)W_1^-(s)-(t-s)W_1^-(u)+(s-u)U_1(t)
\nonumber\\
&&+(s-u)\Big\{ 
-16\int_\mpd^\cut {ds'}{3 Im f_1^-(s')\over(s'-\mmd)(s'-\mpd)}
+G_b^-(2)+t G_b^-(3)+t^2 G_b^-(4)
\nonumber\\
&&+H_b^-(2)+(2\Skpi-t) H_b^-(3)+[ (2\Skpi-t)^2-b] H_b^-(4) 
+(b-us)H^-(4)\Big\}
\nonumber\\
&&+O(p^8)\ .
\ena 
On the right-hand sides of eqs. \rf{drplus}\rf{drmin} the dependence
on the cutoff $\Lambda$ must cancel: we have verified that it does, up to
$O(p^8)$ terms. The dependence upon the parameter $b$ must also cancel. This
gives rise to constraints among the $\pi K$ and $\pi\pi\to K\overline K$
amplitudes and their derivatives which we have not explored.

\noindent{\large\bf 3. Chiral representation and sum rules}

\noindent{\bf 3.1 Chiral representation at $O(p^4)$}

First, let us recall, that at the leading chiral order, $O(p^2)$,
one has 
\be
F^{1\over2}(s,t)={1\over4f_\pi^2}(4s+3t-4\Skpi),\quad
F^{3\over2}(s,t)={1\over4f_\pi^2}(-2s +2\Skpi)
\en
or
\be
F^+(s,t)={1\over4f^2_\pi} t,\quad
F^-(s,t)={1\over4f^2_\pi}(s-u)\ .
\en
The corresponding $\pi K$ partial waves, are, first for $l=0$ 
\bea\lbl{f0p2}
f_0^{1\over2}(s)={1\over128\pi f_\pi^2}
\left(5s-2\Skpi-{3\Del^2\over s}\right),\quad
f_0^{3\over2}(s)={1\over64\pi  f_\pi^2}
(-2s+2\Skpi)\ ,
\ena
then for $l=1$
\bea\lbl{f1p2}
f_1^{1\over2}(s)={1\over128\pi f_\pi^2}
(s-2\Skpi+{\Del^2\over s}),\quad
f_1^{3\over2}(s)=0,
\ena
while the partial waves for $l\ge2$ vanish at this order. In the
$\pi\pi\to K\overline K$ channel, the $l=0$ and $l=1$ partial waves are
\be
g_0^0(t)={\sqrt3 t\over 64\pi f_\pi^2},\quad
g^1_1(t)= {\sqrt 2\over 48\pi f_\pi^2}\ .
\en
At the NLO order, according to the discussion above, the 
$\pi K$ amplitudes must have the following form,
\bea\lbl{chplus}
&&F^+(s,t)=
\left[\overline W_0^+(s)+(t-u)\overline W_1^+(s) +(s\leftrightarrow u)\right]
+\overline U_0(t)\nonumber\\
&&+\lambda_1^+ t^2 +\lambda_2^+ (s-u)^2 +\beta^+ t +\alpha^+
\ena
and
\bea\lbl{chmin}
&&F^-(s,t)=
\left[\overline W_0^-(s)+(t-u)\overline W_1^-(s) -(s\leftrightarrow u)\right]
+(s-u) \overline U_1(t)\nonumber\\
&&+(s-u)( \lambda_1^- t +\beta^-)\ .
\ena
Indeed, the calculation was performed in ChPT at $O(p^4)$ in ref. \cite{bkm},
and it is not difficult to recast their result in the above form. We display
the explicit expressions below, which will be used in the derivation
of the sum rules. The $W$ functions receive contributions from $\pi K$
and $\eta K$ intermediate states,
\be
\overline W^\pm_l(s)={1\over64 f^4_\pi}\left(
W_{l,\pi K}^\pm(s)+ W_{l,\eta K}^\pm(s)\right)\ .
\en
For the $W^+_{0,PQ}$ functions, one obtains from \cite{bkm}
\bea\lbl{Wp4}
&&W^+_{0,\pi K}=
\Big[ 19s^2-28s\Skpi+12\Skpi^2-9\Del^2+{2\Del^2\Skpi\over s} 
+{4\Dkpi^4\over s^2}\Big ]\Jpik\nonumber\\
&&\phantom{W^+_{0,\pi K}=}
-{4\Dkpi^4\over s}\JJpik\nonumber\\
&&W^+_{0,\eta K}=
\Bigg[ 3s^2-4s\Skpi+{4\over3}\Skpi^2+\Del^2+6\Dkpi\Dketa
-{2\Dkpi\over s}(\Dkpi\Sketa\nonumber\\
&&\phantom{W^+_{0,\eta K}=}
+2\Dketa\Skpi)+{4\Dkpi^2\Dketa^2\over s^2}\Bigg]\Jketa
-{4\Dkpi^2\Dketa^2\over s}\JJketa\ .
\ena
In these expressions, 
$\bar J_{PQ}(s)$ is the standard one-loop function \cite{gl85} which has
the following dispersive representation
\be
\bar J_{PQ}(s)={s\over 16\pi^2}\int_{(m_P+m_Q)^2}^\infty
ds'{\sqrt{\lambda_{PQ}(s')}
\over (s')^2(s'-s)}\ ,
\en
with 
\be
\lambda_{PQ}(s')=(s'-(m_P+m_Q)^2)(s'-(m_P-m_Q)^2)\ .
\en
Now the $W^-_{0,PQ}$ functions are
\bea
&&W^-_{0,\pi K}=W^+_{0,\pi K}-16(s-\Skpi)^2
\nonumber\\
&&W^-_{0,\eta K}=W^+_{0,\eta K}
\ena
The last equality holds because the $\eta K$ state  
has isospin $I=1/2$. Also 
$Im f_1^{3\over2}$ vanishes at $O(p^4)$ (and also at $O(p^6)$) and
consequently,
\be
\overline W^+_1(s)=\overline W^-_1(s)\ .
\en
The expression for the $W^+_{1,PQ}$ components is the same for
$PQ=\pi K$ or $\eta K$ and is given by
\be
W^+_{1,PQ}=(s-2\Sigma_{PQ}+{\Delta^2_{PQ}\over s})\bar J_{PQ}(s)
-4\Delta^2_{PQ} \bar J'_{PQ}(0)\ .
\en
Finally, the $\overline U_l$ functions at $O(p^4)$ are
\bea
&&16f^4_\pi\, \overline U_0(t)=2t(2t-m^2_\pi)\Jpipi+3t^2\Jkk
+2m^2_\pi(t-{8\over9}m^2_K)\bar J_{\eta\eta}(t)\ ,
\nonumber\\
&&48f^4_\pi\, \overline U_1(t)=2(t-4m^2_\pi)\Jpipi+(t-4m^2_K)\Jkk\ .
\ena
The imaginary 
parts of the $\overline W$ and $\overline U$ functions at $O(p^4)$ can be recovered
from the definitions eqs. \rf{wplusdef}\rf{wmindef} in 
terms of the imaginary parts of $f_l^\pm(s)$,
$g_0^0(s)$ and $g_1^1(s)$ and using unitarity to relate the latter
to the $l=0,1$ amplitudes 
$\pi K\to \pi K,\ \eta K$ and  $\pi \pi\to \pi \pi,\ K \overline K,\ 
\eta \eta$ computed at $O(p^2)$. For instance, unitarity gives
\be
{1\over64 f_\pi^4} Im W^+_{0,\pi K}(s)=
16\pi {\sqrt{\lambda_{\pi K}(s)}\over s}\left[
{1\over3}\vert f_0^{1\over2}(s)\vert^2 +
{2\over3}\vert f_0^{3\over2}(s)\vert^2 +
{\Dkpi\over \lambda_{\pi K}(s)}\vert f_1^{1\over2}(s)\vert^2 
\right]\ ,
\en
and using the $O(p^2)$ expressions \rf{f0p2}\rf{f1p2} for the partial waves
one recovers the same imaginary part as in eq. \rf{Wp4}.

The separation in eqs. \rf{chplus}\rf{chmin} into a polynomial
part and a part with cut-analytic functions is arbitrary: we have only
required that each piece be scale independent 
(a different choice was made in ref. \cite{ab}) and finite.  The coefficients
of the polynomials are simple linear functions of the coupling-constants
$L_i(\mu)$. Using the following notation,
\be
L_P=\log{m^2_P\over\mu^2},\ R_{PQ}={m^2_P\over m^2_P-m^2_Q}
\log{m^2_P\over m^2_Q}
\en
and the result of ref. \cite{bkm}
one finds for the coefficients entering the $F^-$ amplitude at $O(p^4)$
\bea\lbl{coefmin}
&&f^2_\pi \beta^-= {1\over 4} +{2m^2_\pi\over f^2_\pi}\left[
L_5 -{1\over512\pi^2}(6L_K+5R_{\pi K}+R_{\eta K})\right]
\nonumber\\
&&f^4_\pi\lambda_1^-=-L_3 +{1\over512\pi^2}
\left[ -{4\over3}\log{m^2_\pi\over m^2_K}
+R_{\eta K}+ R_{\pi K} \right]\ .
\ena
The coefficients entering the $F^+$ amplitude, then, have
the following expression in terms of the $L_i$'s,
\bea\lbl{coefplus}
&&\alpha^+={8m^2_\pi m^2_K\over f^4_\pi}\Big\{
4L_1+L_3-4L_4-L_5+4L_6+2L_8\nonumber\\
&&\phantom{\alpha^+={8m^2_\pi m^2_K\over f^4_\pi}}
+{1\over512\pi^2}\left[
{7\over9}L_\eta-L_K-R_{\pi K}+{1\over3}R_{K\eta}-{2\over9}\right]\Big\}
\nonumber\\
&&\beta^+=\beta^-
+{8(m^2_\pi+m^2_K)\over f^4_\pi}\left[
-2L_1-{1\over2}L_3+L_4+{1\over512\pi^2}\left(L_K+R_{\pi K}+{1\over3}\right)
\right]
\nonumber\\
&&\quad +{m^2_\pi\over128\pi^2 f^4_\pi}\log{m^2_\pi\over m^2_\eta}
\nonumber\\
&&f^4_\pi\lambda_1^+=8L_1+2L_2+{5\over2}L_3+{1\over512\pi^2}
\left[-8L_\pi-10L_K-4R_{\pi K}-15\right]
\nonumber\\
&&f^4_\pi\lambda_2^+=2L_2+{1\over2}L_3+{1\over512\pi^2}
\left[-6L_K-5R_{\pi K}-R_{\eta K}+{1\over3}\right]\ .
\ena
This completes the rewriting of the chiral formulas of ref. \cite{bkm}
in a form which allows easy matching with the dispersive 
representations. This matching generates a number of sum rules. 
For $F^-$,  the dispersive formula has no subtraction constant, which
implies that the two coefficients $\beta^-$ and $\lambda_1^-$ can be expressed
as sum rules. For $F^+$, one subtraction constant
remains and this implies 
that the three coefficients $\beta^+$, $\lambda_1^+$, $\lambda_2^+$ are
expressible as sum rules while the fourth one, $\alpha^+$, remains
undetermined in this approach. 
Using eqs. \rf{coefmin}\rf{coefplus} it is then easy to generate sum rule
expressions for the  $L_i$ coupling constants, which are given in terms of
$\beta^\pm$, $\lambda_i^\pm$ as simple linear combinations. For instance,
$L_1$, $L_2$ are given by
\bea
&&L_1={f^4_\pi\over8}(\lambda_1^+-\lambda_2^++2\lambda_1^-)
-{1\over512\pi^2}\left(-{4\over3}L_\pi-{1\over6}L_K+{3\over8}R_{\pi K}
+{3\over8}R_{\eta K}-{23\over12}\right)
\nonumber\\
&&L_2={f^4_\pi\over4}(2\lambda_2^++\lambda_1^-)
-{1\over512\pi^2}\left(-{1\over3}L_\pi-{8\over3}L_K-{9\over4}R_{\pi K}
-{1\over4}R_{\eta K}+{1\over6}\right)\ .
\ena
while $L_3$ is immediately given in terms of $\lambda_1^-$. 
The coupling $L_4$, finally, is obtained from the following combination
\bea\lbl{L4}
&&L_4={f^4_\pi\over8}\left({\beta^+-\beta^-\over m^2_K+m^2_\pi}+
2(\lambda_1^+-\lambda_2^+) \right)
\nonumber\\
&&\quad 
-{1\over512\pi^2}\left(-2L_\pi+{5\over4}R_{\pi K}+{1\over4}R_{\eta K}
-{7\over2}
+{m^2_\pi\over2(m^2_\pi+m^2_K)}\log{m^2_\pi\over m^2_\eta}\right)\ .
\ena
We observe that while the coupling $L_5$ is present in the expression for
$\beta^-$, it appears multiplied by $m^2_\pi$ (not $m^2_K$) 
and  thus makes a very small
correction to the leading $O(p^2)$ contribution. 

\noindent{\bf 3.2 Sum rules}

The dispersive representation of the $\pi K$ amplitudes eqs. \rf{drplus}
\rf{drmin} contains one arbitrary parameter, while the polynomial part
of the chiral representation eqs. \rf{chplus}\rf{chmin} at $O(p^4)$, 
contains six coefficients: comparing the two representations should
yield five sum rules for these coefficients which will translate,
in principle,
using expressions  \rf{coefplus}\rf{coefmin} into sum rules for the five
coupling constants $L_i(\mu),\ i=1,5$.
The explicit form of the sum rules are obtained by noting that 
differences like 
$W^+_0(s)-\overline W^+_0(s)$, in which $\overline W^+_0$ is 
computed to $O(p^4)$ accuracy,
are analytic up to $O(p^6)$ contributions,
\be
Im(W^+_0(s)-\overline W^+_0(s))=O(p^6)\ .
\en 
Therefore, up to $O(p^6)$ terms, we can expand these differences
as polynomials,
\bea\lbl{Aldef}
&& W_l^\pm(z)- \overline W_l^\pm(z)=A_l^\pm + B_l^\pm z +C_l^\pm z^2
\nonumber\\
&& U_l(z)    - \overline U_l(z)    =u_l     + v_l     z +w_l     z^2\ ,
\ena
for $l=0,1$.    We also introduce
\be
\hat A^\pm_1=16\int_\mpd^\cut{3Im f_1^\pm(s')\,ds'\over
(s'-\mpd)(s'-\mmd)}\ .
\en
The  coefficients $A^\pm_l,\ B^\pm_l$ etc... are given as inverse moments 
of the imaginary parts of the $\pi K$ and $\pi\pi\to K\overline K$ $S$ and $P$
waves, integrated between the threshold and $\cut$,
with the chiral part being subtracted.
Together with the integrals \rf{heplus}\rf{hemin} 
over the high-energy region, $[\cut,\infty]$, they form the building blocks 
of the sum rules.

Equating the chiral and the dispersive expressions, taking into account
eqs. \rf{Aldef}, we finally obtain the following sum rule formulas for the
polynomial coefficients in the 
$O(p^4)$ chiral representation \rf{chplus}\rf{chmin}
\bea\lbl{sr}
&&\beta^-=-\hat A_1^- +A_1^-+B_0^-  +u_1 +2\Skpi C_0^-
+G_b^-(2)+H_b^-(2)+2\Skpi H_b^-(3)
\nonumber\\
&&\beta^+=\hat A_1^+  +3A_1^+ -B_0^++v_0 +2\Skpi(2B_1^+-C_0^+)
+G_b^+(2)
\nonumber\\
&&\quad\quad -H_b^+(2)+2\Skpi(H^+(3)-2H_b^+(3))
\nonumber\\
&&\lambda_1^+=-{3\over2}B_1^++{1\over2}C_0^++w_0
+G_b^+(3)-{1\over2}H^+(3)+H_b^+(3)
\nonumber\\
&&\lambda_2^+={1\over2}B_1^++{1\over2}C_0^++{1\over2}H^+(3)
\nonumber\\
&&\lambda_1^-=B_1^--C_0^-+v_1
+G_b^-(3)-H_b^-(3)\ .
\ena
The derivation and the structure of these sum rules are very similar
to those which were proposed for $\pi\pi$ scattering in ref. \cite{knecht}. 
We now discuss the practical evaluation of these formulas.

\noindent{\large\bf 4. Evaluation of the sum rules}

\noindent{\bf 4.1 $\pi K$ amplitudes}

We will make use of the two most recent high-statistics
$K p$ production experiments, both performed at SLAC,
which have determined  $\pi K$ amplitudes. 
Estabrooks et al. \cite{estabrooks} have considered several charge 
combinations enabling them to determine separately the $I=1/2$ and the
$I=3/2$ combinations.  For the isospin $I=3/2$
it was observed that the P and D waves remain
very small below $\sqrt{s}=2$ GeV: in our calculations we will only
include the S-wave. 
A few years later the $K^-\pi^+\to K^-\pi^+$
amplitude was remeasured in a slightly larger energy range by 
Aston et al. \cite{aston}. For the $I=1/2$ S and P waves, we have 
performed fits of the data of Aston et al. with parametrisations
in terms of Breit-Wigner plus background similar to those used 
in this reference using the $I=3/2$ S-wave
from ref. \cite{estabrooks}. 
In these fits, we have imposed that the
scattering lengths be equal to their values in ChPT \cite{bkm}. Relaxing
this constraint, however, makes very little change in the results. 
For the partial waves $l=2-5$, we have used exactly the same parametrisations
as provided in ref. \cite{aston}. Above $\sqrt{s}=1.5$ GeV, ambiguities
arise in the determination of the $S$ and $P$ waves. Estabrooks et al 
find four different solutions and Aston et al. two. It has been pointed
out in ref. \cite{jop} that one of these violate the unitarity bound; 
therefore,we
have used the remaining one. In our sum rules, we note that the contribution
from the $S$ and $P$ waves in this energy region 
becomes negligibly small anyway. Above the energy range covered by these
experiments we use Regge parametrisations which we will discuss in more
detail below.

\noindent{\bf 4.2 $\pi \pi\to K\overline K$ amplitudes}

Let us first discuss the S-wave amplitude, 
\be
g_0^0(t)\equiv |g_0^0(t)| \exp(i\phi_S(t))\ ,
\en
which is a crucial ingredient in the sum rules and is needed for
$t\ge 4m^2_\pi$ while it is measured only in the range
$t\ge 4m^2_K$. Analyticity, as is well known \cite{J&P,J&N,hedegard,ader}, 
is the key to performing
this extrapolation. To start with, the phase
of the amplitude, $\phi_S(t)$, 
may be considered as known in the whole energy region of interest.
Firstly, in the  region where $\pi\pi$ scattering is elastic, $\phi_S$ 
is identical to the $\pi\pi$ phase shift (modulo $\pi$)
from Watson's theorem. It is now well 
established that, to a very good approximation, 
the domain where $\pi\pi$ scattering is effectively elastic extends up to 
the $K \overline K$ threshold (see e.g. \cite{mms}).
Above this point, $\phi_S(t)$  has been measured in
experiments, we will use the two most recent ones: 
Cohen et al. \cite{cohen}
(who considered $K^+ K^-$ production) 
and Etkin et al. \cite{etkin,longacre}
(who considered $K^0_S K^0_S$). These data are shown in 
Fig. 1 together with the curves which will be used in the calculations. 
\begin{figure}[thb]
\leavevmode
\begin{center}
\includegraphics[width=14cm]{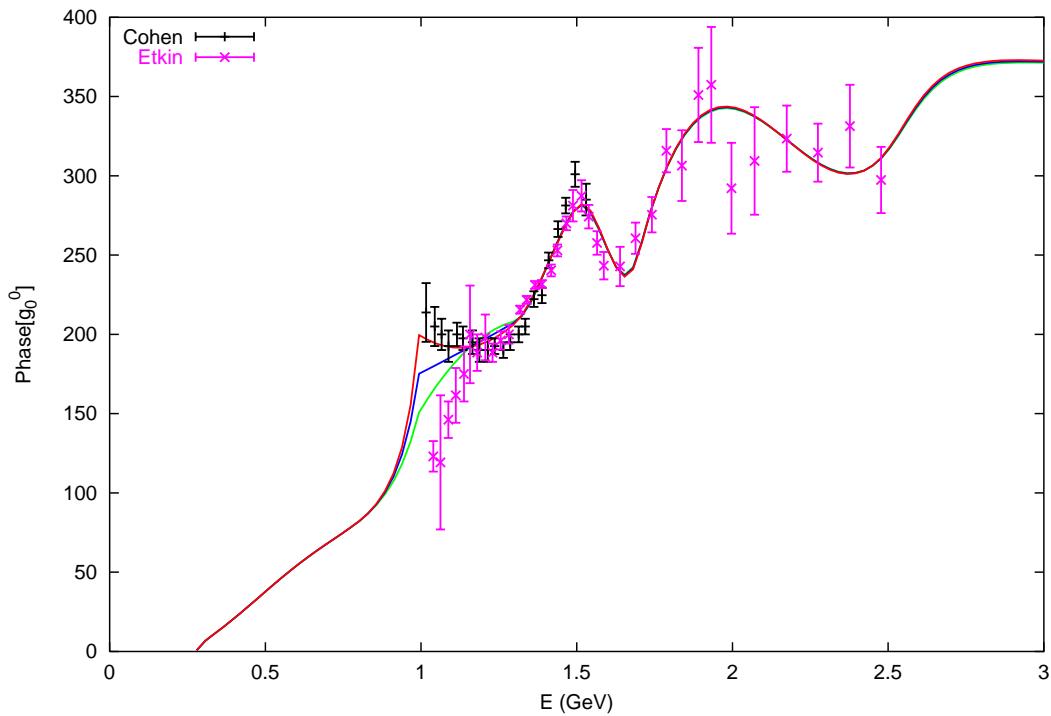}
\end{center}
\caption{\sl Phase of the $\pi\pi\to K\overline K$ $l=0$ amplitude. 
The curves are
as described in the text. The three curves from bottom to top correspond
to three increasing input values for the phase at 
the $ K\overline K$ threshold,
$\phi_S(t_0)=150^\circ,\ 175^\circ,\ 200^\circ$.}
\label{Figure 1}
\end{figure}
One observes that the two data sets are compatible except very close to
the $K \overline K$ threshold. (It must be recalled here that 
experiments actually
measure  $S-D$ interference and thus determine only the difference 
$\phi_S -\phi_D$. In treating the data of 
Etkin et al. we have used the $D$-wave model of ref. \cite{amp} rather 
than that used in the original paper \cite{etkin}.) 
In the energy region $\sqrt{t}> 2m_K$, we fit the combined set of
data with piecewise polynomials.
In performing the fit, we have excluded the small energy region
where the two data sets are incompatible and we have instead fixed
the value of the phase at threshold $\phi_S(4m^2_K)$, which we have
allowed to vary between $150^\circ$ and $220^\circ$. 
This may seem like a wide range and one could think of making use
of the equality between $\phi_S$ and the $\pi\pi$ phase-shift at
the threshold to improve on that. However, the  available $\pi\pi$ data points
closest to the $K\overline K$ threshold have large error bars.
In order to get exactly at the threshold, one needs to perform a fit
and the result is uncertain because the $\pi\pi$ phase-shift 
varies extremely rapidly in this region \cite{mms}. One ends up
with the same range of values as we have chosen.
In the energy range $\sqrt{t}\le 0.8$ GeV, the curve in Fig. 1 
is the result from
the recent Roy equations analysis of ref. \cite{acgl}, with $a_0^0=0.22$, 
$a_0^2=-0.0444$. Because of the new $K_{l4}$ data \cite{newkl4} there is now
a rather small uncertainty on the curve in this energy region \cite{cgl}.
In the energy region between $0.8$ GeV and the $K \overline K$ threshold we use
the simple interpolation formula,
\be
\phi_S(t)=\alpha +{\beta\over\sqrt{t}-E_1}
\en 
in which the three parameters are determined from imposing continuity
at both ends and continuity of the first derivative at the lower end.

Once the phase is known, determining the
modulus in the region $[4m^2_\pi, 4m^2_K]$ is a   
standard Muskhelishvili-Omn\`es \cite{muskh,omnes} 
problem because $g_0^0$ satisfies the following integral equation
\be\lbl{g00dr}
g_0^0(t)={t\over\pi}\int_{4m^2_\pi}^\infty {Im g_0^0(t')dt'
\over t' (t'-t)} +g_0^0(0) +\Delta(t)
\en
where $\Delta(t)$ has only a left-hand cut. $\Delta(t)$ can be expressed 
explicitly in terms of $\pi K$ partial wave amplitudes 
and  $\pi \pi\to K \overline K$ partial waves with $l\ge 2$,
\be\lbl{delta}
\Delta(t)=\sum_{l} \int_{m^2_+}^\infty ds' K_{l 0}(s',t) Im f_l^+(s')
+\sum_{l\ge2, even} \int_{4m^2_\pi}^\infty dt' G_{l 0}(t',t) 
(q_\pi q_K)^l Im g_l^0(t')\ .
\en
For $t\lapprox1$ GeV$^2$, $\Delta(t)$ is dominated by the $\pi K$ $S$ and
$P$ waves and eq. \rf{g00dr} is one component of a system of 
Roy-type equations \cite{roy,steiner}. This system was expressed
in ref. \cite{ab} in terms of the two scattering lengths $a^{1/2}_0$, 
$a^{3/2}_0$. As we do not attempt to solve the full system here, we
find it more convenient to use $g_0^0(0)$ as subtraction constant.
We have included $D$ waves as well into the calculation in order to
extend the validity of 
the evaluation of $\Delta(t)$ somewhat above one GeV. 
The kernels needed in eq. \rf{delta} are
\bea
&&K_{00}(s',t)=I_0(s',t)-I_0(s',0),\ 
\nonumber\\
&&{\rm with}\ I_0(s',t)=
{4\over\sqrt{(t-4m^2_\pi)(t-4m^2_K)}}{\rm Arcth}
{\sqrt{(t-4m^2_\pi)(t-4m^2_K)}\over 2s'-2\Skpi+t}
\nonumber\\
&&K_{10}(s',t)=3\left[ I_0(s',t)\left(1+{2s't\over \lambda_{\pi K}(s')}\right)
-{2t\over \lambda_{\pi K}(s')}-I_0(s',0)\right]
\nonumber\\
&&K_{20}(s',t)=5\Bigg[ I_0(s',t)\left(1+{6s't\over \lambda_{\pi K}(s')}
+{6(s't)^2\over \lambda^2_{\pi K}(s')}\right)
\nonumber\\
&&\quad +{6t\over \lambda^2_{\pi K}(s')}\left( s'(-2s'+2\Skpi-t)+{1\over6}
(t-4m^2_\pi)(t-4m^2_K)\right)-I_0(s',0)\Bigg]
\nonumber\\
&&G_{20}(t',t)={16\times5\over\sqrt3}
{t'(t'+t-4\Skpi)\over t'(t'-4m^2_\pi)(t'-4m^2_K)}
\ .
\ena
The various contributions  and the result for 
$\Delta(t)$ are displayed in Fig. 2. 
\begin{figure}[thb]
\leavevmode
\begin{center}
\includegraphics[width=14cm]{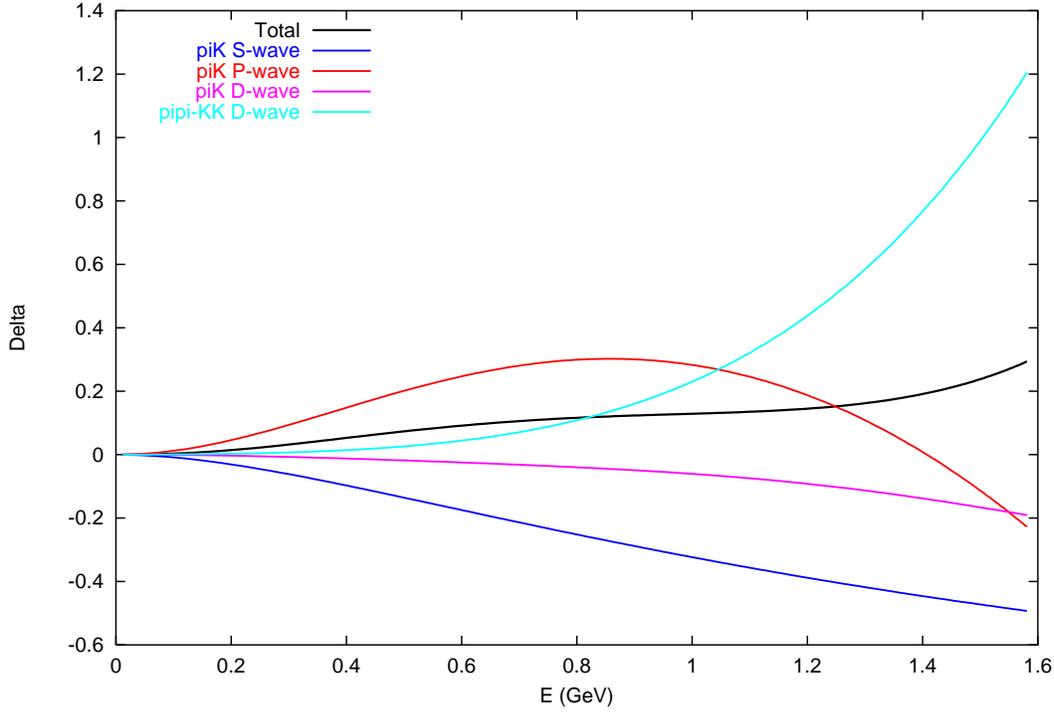}
\end{center}
\caption{\sl 
Left-hand cut function, $\Delta(t)$, and its various contributions.
}
\label{Figure 2}
\end{figure}

In order to solve equation \rf{g00dr}
we first construct the Omn\`es function over the range $[4m^2_\pi, t_0]$,
with $t_0=4m^2_K$
\be\lbl{omdef}
\Omega(t)=\exp\left[ {t\over\pi} \int_{4m^2_\pi}^{t_0} {\phi_S(t')dt'
\over t'(t'-t)}\right]\equiv
\Omega_R(t)\exp[i \phi_S(t)\theta(t-4m^2_\pi)\theta(t_0-t)]
\en
where $\Omega_R(t)$ is real. When $t$ approaches the $K \overline K$ threshold,
$\Omega_R(t)$ has the following behaviour,
\be
\lim_{t\to 4m^2_K} \Omega_R(t)\sim \vert t-4m^2_K\vert^
{\phi_S(4m^2_K)\over\pi}\ .
\en
Therefore, two cases must be considered
depending whether $\phi_S(4m^2_K)$ is smaller or larger than $\pi$.
Let us first consider the case 
\be\lbl{cond}
\phi_S(4m^2_K)\le\pi\ .
\en 
The solution to 
eq. \rf{g00dr} is obtained by introducing the function
\be\lbl{ft}
f(t)={1\over \Omega(t)}(g_0^0(t)-\Delta(t))
\en
and noting that it is analytic except for a right-hand
cut. It can thus be expressed as a dispersion relation
which is
defined up to a polynomial which depends on the behaviour at
infinity of $f(t)$ \cite{muskh}. We will assume that $f(t)$ is bounded
at infinity by a polynomial of degree one and thus, $g_0^0$ can be
expressed in terms of two subtraction constants,
\bea\lbl{g00sol}
&&g_0^0(t)=\Delta(t) +\Omega(t)\Big[ \alpha_0 +\beta_0 t
+{t^2\over\pi}\int_{4m^2_\pi}^{4m^2_K}dt' { \Delta(t')\sin\phi_S(t')
\over\Omega_R(t') (t')^2(t'-t)}\nonumber\\
&&\quad +{t^2\over\pi}\int_{4m^2_K}^\infty dt' {\vert g_0^0(t')\vert
\sin\phi_S(t')\over \Omega_R(t') (t')^2(t'-t)}\Big]\ .
\ena
One observes that the integrals converge at $t'=4m^2_K$ if the
condition \rf{cond} is satisfied. A small calculation shows that at
$t=4m^2_K$ the condition $(g_0^0)_{output}=(g_0^0)_{input}$ is automatically
satisfied in eq. \rf{g00sol}. 
Concerning the parameters $\alpha_0,\ \beta_0$ it is not
difficult to see that for the purpose of using $g_0^0$ 
with $O(p^4)$ precision it is consistent to use the values of 
$\alpha_0,\ \beta_0$  with  $O(p^2)$ precision, i.e.
\be
\alpha_0=0\quad \beta_0={\sqrt3\over64\pi f^2_\pi}-\Delta'(0)
\en
and $g_0^0$ gets fully determined (the value of the derivative $\Delta'(0)$
is determined numerically to be $\Delta'(0)\simeq 0.256$ GeV$^{-2}$). 
The influence of $\Delta(t)$ is illustrated
in Fig.3 which compares the full solution from eq. \rf{g00sol} to the
solution with $\Delta$ set equal to zero. 
\begin{figure}[thb]
\leavevmode
\begin{center}
\includegraphics[width=14cm]{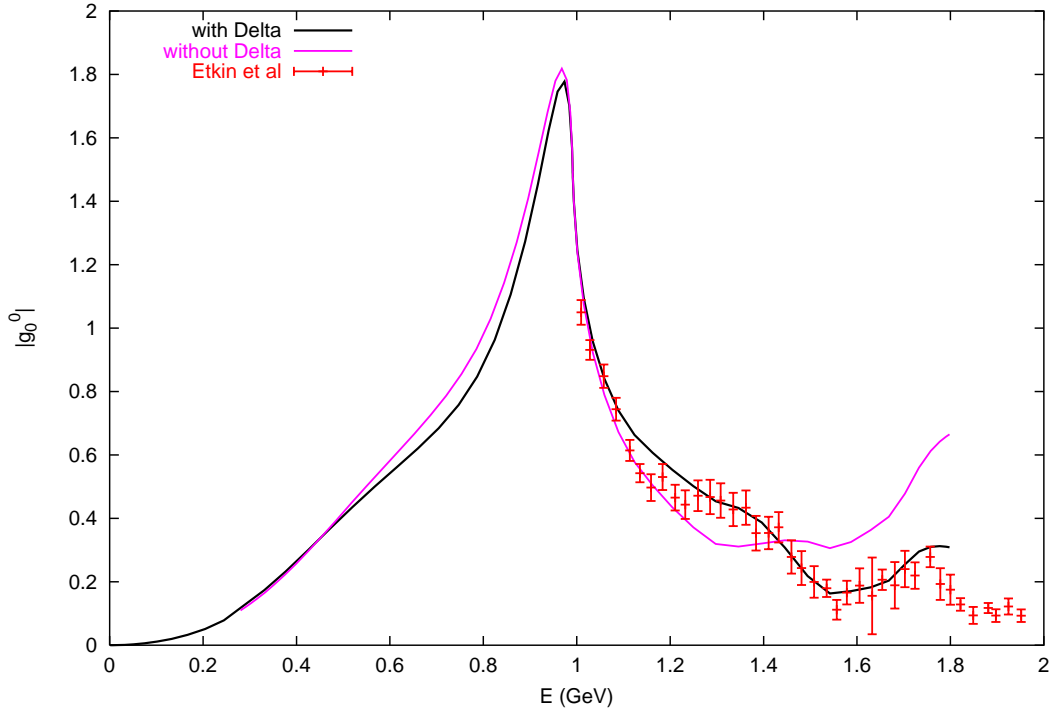}
\end{center}
\caption{\sl Solutions of eq. \rf{g00dr} with $\Delta(t)$ included
and $\Delta(t)$ ignored.
}
\label{Figure 3}
\end{figure}
One sees that in the 
energy region where we really need to use eq. \rf{g00sol}, i.e.
below the $K \overline K$ threshold $\Delta(t)$ 
actually has a rather small effect.
The solution is essentially controlled from the chiral constraints
at $t=0$ and the experimental input at $t\ge 4m^2_K$. 
Above the $K \overline K$  threshold, the agreement of the 
experimental data with
the output from eq. \rf{g00sol} seems much improved if $\Delta(t)$
is included. In the case where 
\be\lbl{condb}
\phi_S(4m^2_K)>\pi\ ,
\en 
we need only modify the definition of $f(t)$ (eq. \rf{ft}) to
\be
f(t)={t-4m^2_K\over \Omega(t)}(g_0^0(t)-\Delta(t))
\en
and make the corresponding change in the preceeding formulas. 
Because of the extra factor of $t$ one needs to introduce
one more subtraction, and an additional parameter, $\gamma_0$ 
appears in the solution: we simply fix $\gamma_0$ so that input-output
agreement is retained in the physical region when the condition 
 \rf{condb} holds. 

There is a subtlety in the above calculation which must be discussed. 
Clearly, because of the singular behaviour of the term $1/\Omega_R(t')$
in the integrand at the $K \overline K$ threshold the result will be rather
sensitive to the value of $\vert g_0^0(t')\vert$ in this region and, 
in particular, to its value exactly at the threshold, while experimental
information starts slightly above the $K \overline K$ threshold. A simple 
solution is to use $t_0$ slightly larger than $4m^2_K$ in eq. \rf{omdef}.
We have done so and found good stability of the result.
We have also used the
following simple method which, at the same time, provides
an alternative extrapolation method in the whole region of interest. We
first construct an Omn\`es function over a region $[4m^2_\pi, t_1]$ with
$t_1>> 4m^2_K$,
\be
\Omega_1(t)=\exp\left[ {t\over\pi} \int_{4m^2_\pi}^{t_1} 
{\phi_S(t')dt'\over t'(t'-t)}\right]\ ,
\en
and then consider the function
\be\lbl{V00}
V_0^0(t)={1\over \Omega_1(t)} g_0^0(t)\ .
\en
The function $V_0^0$ is analytic with a left-hand cut, and a 
right-hand cut which only starts at $t=t_1$. Therefore, $V_0^0$ is expected
to be a smooth function in the region $[0,t_1]$ and we can use approximations
by polynomials there. 
In practice, we used fourth order polynomials,
\be\lbl{V00pol}
V_0^0(t)=\alpha_0 +\tilde\beta_0 t +\tilde\gamma_0 t^2 +\delta_0 t^3,\quad
0\le t\le t_{fit}< t_1\ .
\en
The first two parameters are determined from ChPT as above and we fit
the remaining two to values of $V_0^0$ determined
from the data above the $K \overline K$ threshold. The energy range in which
the fit is performed $t\le t_{fit}$ cannot, of course, be made two large
otherwise higher order polynomials would be needed. Fig. 4 shows 
two different fits and illustrates that this procedure, while simple, is
also quite stable. 
\begin{figure}[thb]
\leavevmode
\begin{center}
\includegraphics[width=14cm]{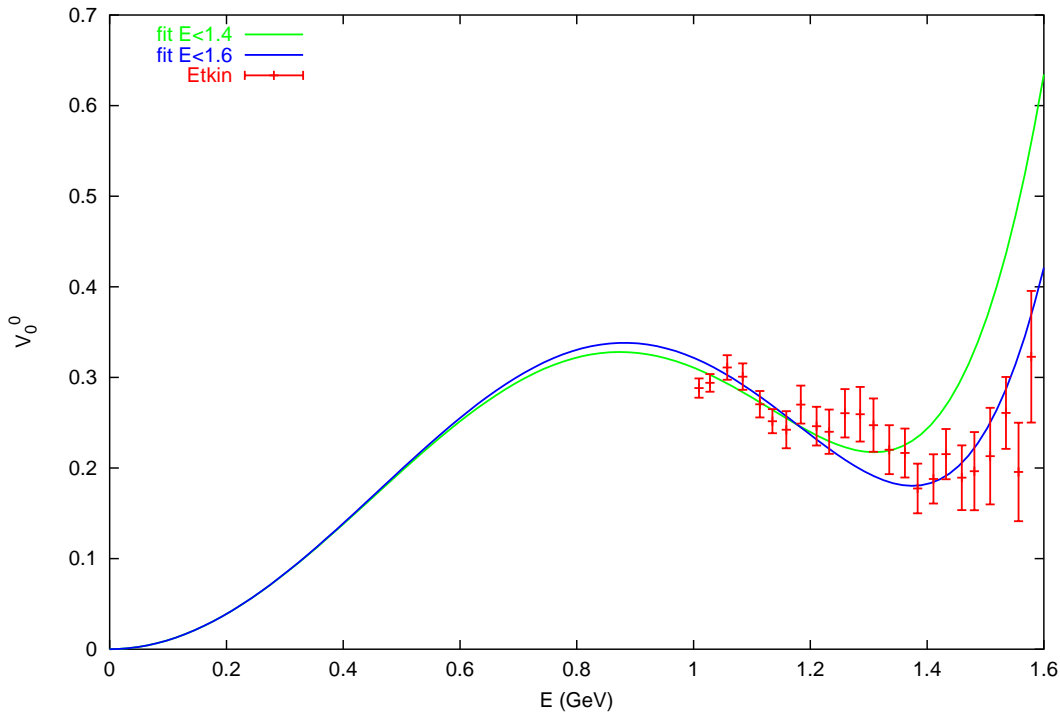}
\end{center}
\caption{\sl Polynomial fits of $g_0^0$ with the right-hand cut 
(partly) removed (see eq. \rf{V00}). 
}
\label{Figure 4}
\end{figure}
This procedure allows one to determine
the value of $\vert g_0^0(4m^2_K)\vert$ (which is thus correlated with
the value of $\phi_S (4m^2_K)$) and can be used for correctly computing
the integrals above  \rf{g00sol}. We can also use the polynomial approximation
to $V_0^0$ to extrapolate $g_0^0$ below the 
$K \overline K$ threshold (note that 
this method requires no knowledge of the left-hand cut and no assumption
concerning asymptotic behaviour). We found that the two methods of 
extrapolation are in very good agreement.

The solution for $g_0^0$ has a rather strong dependence on the value
of the phase $\phi_S$ at the $K \overline K$ threshold as is shown
in fig. 5: the larger $\phi_S(4m^2_K)$ the higher is the corresponding
$f_0(980)$ resonance peak. 
\begin{figure}[thb]
\leavevmode
\begin{center}
\includegraphics[width=14cm]{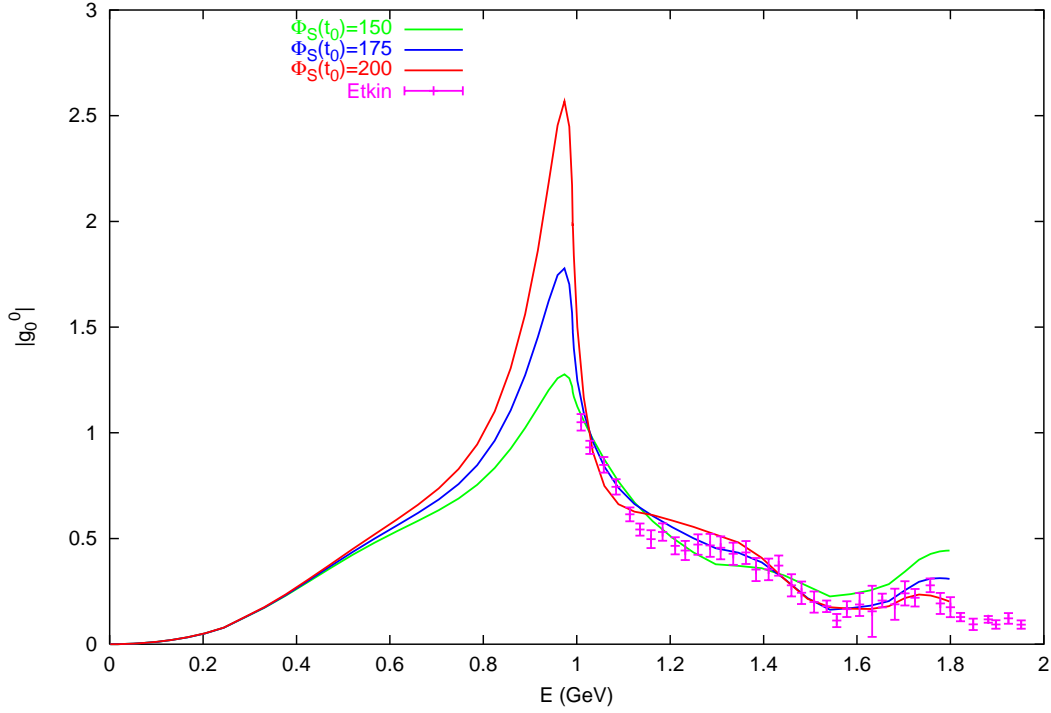}
\end{center}
\caption{\sl Comparison of several solutions of eq. \rf{g00dr} for $g_0^0(t)$
corresponding to different input values of the threshold phase 
$\phi_S(4m^2_K)$.}
\label{Figure 5}
\end{figure}
Another source of uncertainty in this calculation
is the fact that the two available data sets
for $|g_0^0|$, while having small error bars, are not  exactly compatible.
The data of ref. \cite{cohen} lies systematically below the data from
ref. \cite{etkin,longacre}. The corresponding influence in the
$f_0(980)$ peak is shown in fig. 6.
\begin{figure}[thb]
\leavevmode
\begin{center}
\includegraphics[width=14cm]{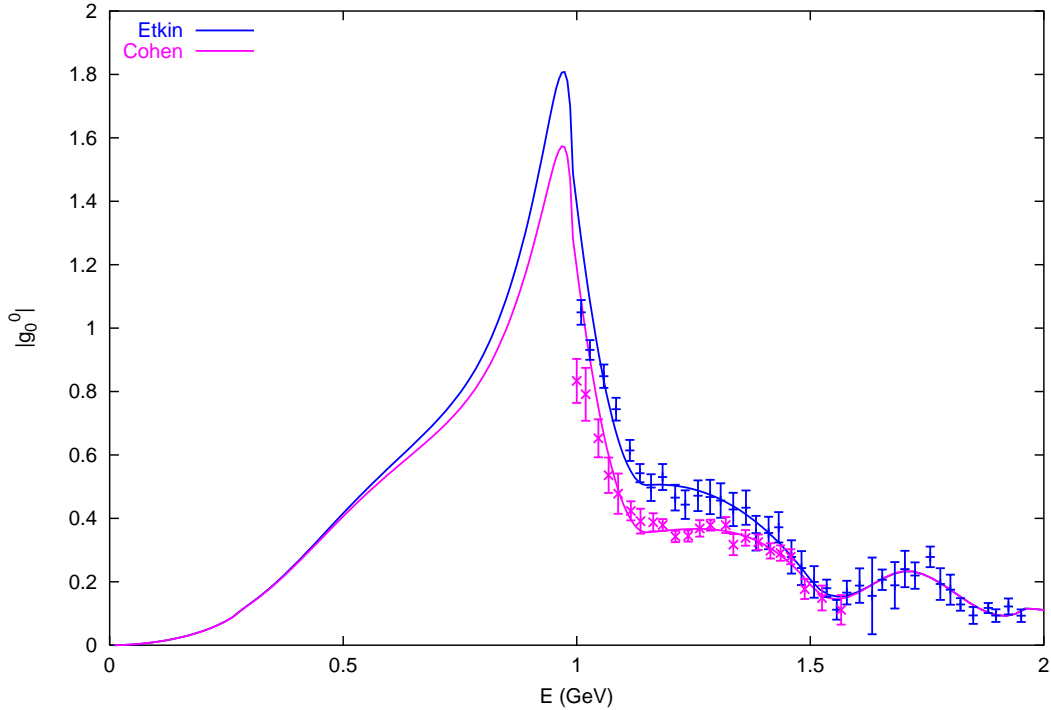}
\end{center}
\caption{\sl Comparison of two solutions for $|g_0^0|$ based on the
two data sets of Cohen et al. \cite{cohen} and 
Etkin et al. \cite{etkin,longacre} respectively. Above the $K \overline K$
threshold the curves are piecewise polynomial fits to the data and below the
threshold they are obtained from eq. \rf{g00sol}.}
\label{Figure 6}
\end{figure}

Let us now turn to the $l=1$ amplitude $g_1^1(t)$. We will again here 
rely on the experimental data from ref. \cite{cohen} above the $K\overline K$
threshold and chiral symmetry at $t=0$. Let us first consider the phase of
$g_1^1$ which we denote by $\phi_P(t)$. In the range 
$\sqrt{t}\le 0.82$ GeV,
$\phi_P(t)$ is equal to the $l=1$ $\pi\pi$ phase-shift and we use
the parametrisation of ref. \cite{acgl} which is 
constrained from the Roy equations. In the range $\sqrt{t}\ge 2m_K$
the measured phase has been shown in ref. \cite{cohen} to be well 
approximated by that of a Breit-Wigner tail of the following form
\be\lbl{g11bw}
g_1^1={m_\rho\sqrt{ \hat G_\pi(t) \hat G_K(t) }/2
\over t-m^2_\rho-i m_\rho( G_\pi(t)+G_K(t))}
\en
with
\be
\hat G_\pi(t)={m_\rho\Gamma_\rho \over q_0^3} 
{1+R^2 q_0^2\over 1+R^2 q_\pi^2},\ 
\hat G_K(t)  ={m_\rho \Gamma_\rho\over2 q_0^3} 
{1+R^2 q_0^2\over 1+R^2 q_K^2}
\en
and
\be
G_\pi=(q_\pi^3/\sqrt{t}) \hat G_\pi,\ G_K=  (q_K^3/\sqrt{t})\hat G_K,\ 
q_0^2=m^2_\rho/4-m^2_\pi,\ R=3.5{\ \rm GeV}^{-1}\ . 
\en
The phase from this
formula departs from the measured one above 1.6 GeV but we will ignore this
discrepancy as in this region the $l=1$ amplitude plays little role.
Finally, in the
intermediate region $0.82\ {\rm GeV}\le \sqrt{t}\le 2m_K$
we use the interpolation formula
\be\lbl{phipint}
\tan \phi_P(t)= (a+b t)/(t-m^2_\rho),\ \ 
0.82\ {\rm GeV}\le\sqrt{t}\le 2m_K.
\en
From this phase we can construct an Omn\`es function
\be
\Omega_P(t)=\exp\left[{t\over\pi}\int_{4m^2_\pi}^\infty
{\phi_P(t')dt'\over t'(t'-t)}\right]\ .
\en  
The magnitude of $g_1^1$ remains to be discussed. As in the case of
$g_0^0$ we expect that it can be expressed with a good approximation
as a low order polynomial times the Omn\`es function in the whole energy
range of interest. In fact, earlier studies based on extrapolations
away from the left-hand cut have shown that a constant polynomial 
is sufficient below one GeV \cite{N&O,lang1}. 
With this in mind, we made a fit to the data between 1 and 1.5 GeV with
a polynomial containing a constant term plus a term quadratic
in $t$, 
\be\lbl{g11pol}
g_1^1(t)\simeq\alpha_1 (1 +\beta_1 t^2) \Omega_P(t),\ 
\sqrt{t}\le 1.5{\ \rm GeV}
\en
fixing  $\alpha_1=\sqrt2/48\pi f^2_\pi$ from $O(p^2)$ chiral symmetry. 
A good fit is obtained in this way with $\beta_1=-0.187 {\rm GeV}^{-4}$
such that the quadratic term is
indeed small below 1 GeV (in discussing the errors we will introduce a
linear term as well). The result for $\vert g_1^1\vert$ is shown
in fig. 7.
\begin{figure}[thb]
\leavevmode
\begin{center}
\includegraphics[width=14cm]{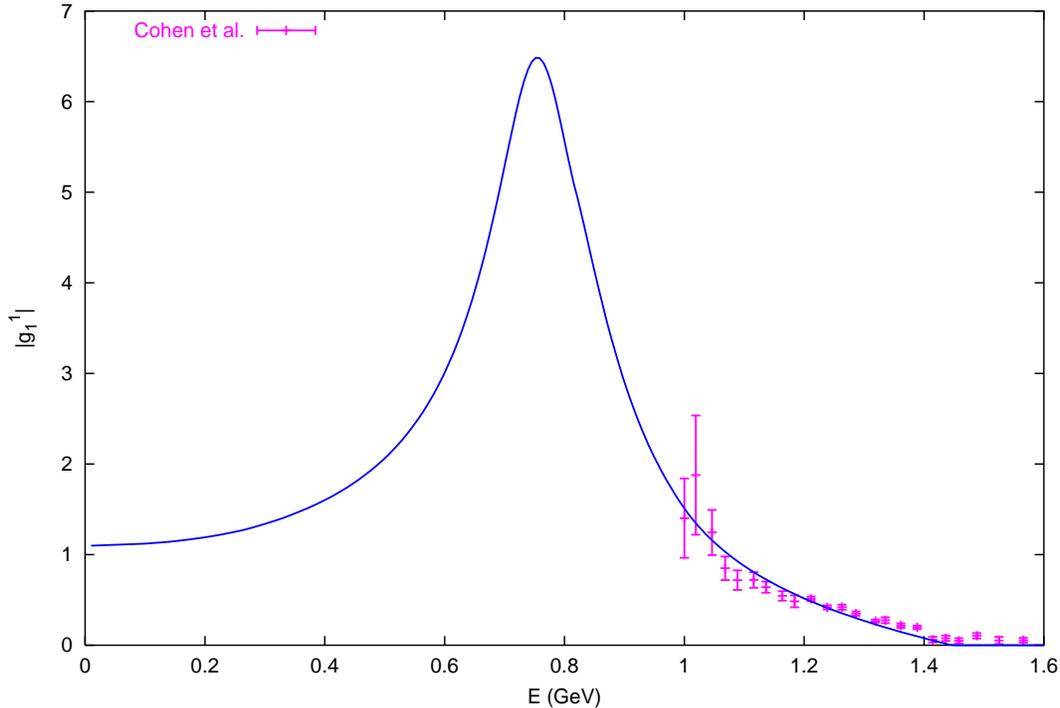}
\end{center}
\caption{\sl Magnitude of the $l=1$ partial wave $g_1^1$ 
(in units of GeV$^{-2}$) from the construction
described in the text, compared to the experimental data \cite{cohen}.
}.
\label{Figure 7}
\end{figure}

We have also included higher partial waves with $l=2,\ 3,\ 4$. For $l=2$ we
include the resonances $f_2(1270)$, $f_2(1425)$, $f_2(1810)$ with
Breit-Wigner functions analogous to eq. \rf{g11bw} and
parameters fitted to the data of ref. \cite{longacre}. For
$l=3$ we include $\rho_3(1690)$ and for $l=4$  the $f_4(2050)$ resonances.  
In both cases we take the  $\pi\pi$ and $K \overline K$ partial
widths  from the PDG \cite{pdg00}.

\noindent{\bf 4.3 Asymptotic region}

Beyond the energy region where the amplitudes are effectively measured
in experiments we can hardly make better than qualitative estimates. For this
purpose we will assume that the resonance region matches to a region where 
Regge behaviour prevails. More specifically, we will consider the 
dual-resonance model for the $\pi K$ amplitude (see e.g. \cite{mms} )
\bea\lbl{veneziano}
&&F^\pm(s,t,u)=-\lambda\left[ V_{K^*\rho}(s,t)\pm V_{K^*\rho}(u,t)\right]
\nonumber\\
&& V_{K^*\rho}(s,t)={\Gamma(1-\alpha_{K^*}(s))\Gamma(1-\alpha_{\rho}(t))
\over\Gamma(1-\alpha_{K^*}(s)-\alpha_{\rho}(t))}\ .
\ena
For the Regge trajectories we take
\be
\alpha_{\rho}(t)=0.475+\alpha_1 t,\quad
\alpha_{K^*}(s) =0.352+\alpha_1 s\ ,\ \alpha_1 =0.882\ {\rm GeV^{-2}}\ ,
\en
and for the parameter $\lambda$ we take $\lambda=1.82$ which realises
an approximate matching to the region known from experiment around
$\sqrt{s}=2$ GeV.
In taking asymptotic limits in formula \rf{veneziano} an $i\epsilon$ 
prescription is understood, for instance, $s\to\infty$ means 
$\vert s\vert\to\infty$ and $s=\vert s\vert\exp(i\epsilon)$. One
then finds the well known Regge behaviour,
\be
Im F^\pm(s,t,u)_{
s\to \infty,\ t\ {\rm fixed}} \sim{\pi \lambda\over\Gamma( \alpha_\rho(t))}
(\alpha_1 s)^{\alpha_\rho(t)}\ .
\en
In the case of $F^+(s,t)$ one needs to include additionally the Pomeron,
\be
Im F^+(s,0)_{Pomeron}={1\over\pi}\sigma s\ ,
\en
in which we take $\sigma=2.5$ mb (see the discussion in ref. \cite{acgl}).
We also need $Im F^\pm(s,t,u)$ in the regime where $t\to\infty$
and $s\to 0$ in the integrals $G_b^\pm(n)$. The model  \rf{veneziano}
gives a Regge behaviour associated with the $K^*$ trajectory,
\be
Im F^\pm(s,t,u)_{t\to\infty, s\ {fixed}}\sim
{\pi\lambda\over\Gamma(\alpha_{K^*}(s))} (\alpha_1 t)^{\alpha_{K^*}(s)}
\ .
\en
Finally, we need $Im F^\pm(s,t,u)$ in a regime where $s\to\infty$ and
$u\to 0$ in the integrals $H_b^\pm(n)$. In this case, the term  
$V_{K^*\rho}(u,t)$ in eq. \rf{veneziano} makes no contribution to the imaginary
part (this, of course, reflects the exact degeneracy of the $K^*$ and
$K^*_2$ trajectories in this model) and the term  $V_{K^*\rho}(s,t)$
becomes exponentially suppressed (this term is the amplitude for the
reaction $\pi^+ K^-\to \pi^+ K^-$ and the corresponding $u-$channel
is $\pi^+ K^+\to \pi^+K^+$, which is exotic). The influence of these
asymptotic contributions can be appreciated from table 1 below.
\begin{table}[hbt]
\begin{center}
\begin{tabular}{|c|c|c|c|c|c|c|}\hline
\ &$H_0^+(2)$&$H_0^+(3)$&$H_b^+(2)$&$H_b^+(3)$&$G_b^+(2)$&$G_b^+(3)$
\\ \hline\hline
$[\cut,4{\rm GeV}^2]$    &8.05&4.66&4.56&3.11&1.26&0.92\\ \hline
$[4{\rm GeV}^2,\infty]$  &8.75&0.72&0   &0   &1.17&0.13\\ \hline\hline
\ &$H_0^-(2)$&$H_0^-(3)$&$H_b^-(2)$&$H_b^-(3)$&$G_b^-(2)$&$G_b^-(3)$
\\ \hline\hline
$[\cut,4{\rm GeV}^2]$    &5.81&    &2.32&1.74&1.76 &1.17\\ \hline
$[4{\rm GeV}^2,\infty]$  &2.18&    &0   &0   &1.25 &0.14\\ \hline
\end{tabular}
\end{center}
\caption{\sl Results for the high-energy integrals (see eqs. \rf{heplus}
\rf{hemin}) in appropriate powers of GeV, showing two different integration
regions.
}
\label{Table 1}
\end{table}

\noindent{\bf 4.4 Results}

Let us first perform some simple checks. Equating two dispersive
representations of $F^-(s,t)$ we obtained eq. \rf{relp}: using this relation
at $t=0$ gives one relation among the building blocks of the sum rules
\be\lbl{cross}
2\hat A_1 +H_0^-(2)= u_1 +G_b^-(2)+ H_b^-(2)
\en
We expect some uncertainty because of the relatively slow convergence in the
integrals $H_0^-(2)$, $G_b^-(2)$ but there is some amount of cancellation
of these effects. Using the phenomenological input as described above,
we obtain,
\be
2\hat A_1 +H_0^-(2)\simeq 33.17\ ,\quad 
u_1 +G_b^-(2)+ H_b^-(2)\simeq 32.07 \ [{\rm \ GeV^{-2}}]. 
\en
Clearly, there is a very reasonable degree of agreement. 
This provides a check on the construction of the 
$\pi\pi\to K\overline K$ P-wave. We have also redone with our input
the calculation of Karabarbounis and Shaw \cite{karab} which
gives the difference in the scattering lengths $a_0^{1/2}-a_0^{3/2}$
\be
m_\pi(a_0^{1/2}-a_0^{3/2})={3m_\pi\over 8\pi(m_\pi+m_K)} F^-(m_+^2,0)
\simeq 0.22\,
\en
to be compared with the result \cite{karab} $m_\pi(a_0^{1/2}-a_0^{3/2})=
0.26 \pm 0.05$. We have a comparable uncertainty due to a large extent
to the asymptotic contributions. A related quantity is the polynomial
parameter $\beta^-$ (eq. \rf{chmin}) 
for which a rather precise value is predicted by the chiral expansion, 
\be
f^2_\pi \beta^- =0.25+0.01+O(p^6)\ ,
\en
where the successive contributions are shown. Using the sum rule expression
\rf{sr} for $\beta^-$ we obtain
\be
f^2_\pi \beta^-\simeq0.24
\en
which is within 10\% of the result from ChPT. The size of the uncertainty
in this calculation is approximately 15\%, so the agreement is satisfactory
but it is not possible to separate the purely $O(p^4)$ part (in other term, 
we have no sum rule for $L_5$).
We note also that the $O(p^4)$ contribution being suppressed,
the $O(p^6)$ one could  be of comparable size.

Let us now discuss the results for the chiral couplings $L_1$, $L_2$,
$L_3$. We recall that these are obtained by first generating sum rules
for the polynomial coefficients $\lambda_1^+$, $\lambda_2^+$, 
$\lambda_1^-$: here the contributions from the asymptotic regions are 
suppressed so we can expect rather good accuracy. Our results
are collected in the last line of table 2,
they complete and update those already given\footnote{
In ref. \cite{ab} a factor $f^2_\pi f^2_K$ was used in eqs. \rf{coefmin}
\rf{coefplus} instead of $f^4_\pi$. Here, we prefer not to include
incomplete parts of the $O(p^6)$ contributions.}
in ref. \cite{ab}.
\begin{table}[hbt]
\begin{center}
\begin{tabular}{|c|c|c|c|l|}\hline
$10^3\times$& $L_1$ & $L_2$ & $L_3$ & $L_3+2L_1$ 
\\ \hline\hline
$\pi\pi\ sr\  O(p^4)$    
           & $-$          &$1.02\pm0.05$&   $-$       &$-2.78\pm0.32$ \\ \hline
$K_{l4}   \phantom{\ sr\ }O(p^4)$
           &$0.46\pm0.23$&$1.49\pm0.23$&$-3.18\pm0.85$&$-2.26\pm0.97$ \\ \hline
$\pi K\  sr\ O(p^4) $    
           &$0.84\pm0.15$&$1.36\pm0.13$&$-3.65\pm0.45$&$-1.97\pm0.34$ \\ \hline
\end{tabular}
\end{center}
\caption{\sl Results from the sum rules
for the chiral couplings $L_1$, $L_2$, 
$L_3$ (multiplied by $10^3$) at the scale $\mu=0.770$ GeV (last line),
compared to the results from $\pi\pi$ sum rules and from the $K_{l4}$
form-factors.}
\label{Table 2}
\end{table}
The way in which the errors quoted in the table are evaluated will be
explained in more details below,  
they do not include any effect from $O(p^6)$ corrections.
The table also shows for comparison the results obtained from $\pi\pi$ 
sum rules (taken from ref. \cite{cgl} in which $O(p^4)$ matching is used) and
the results based on the $K_{l4}$ form-factors also computed at
chiral order $p^4$. 
The numbers quoted in the table are taken from the fit of 
Amoros et al. 
\footnote{We thank P. Talavera for communicating the values of the errors
corresponding to this fit.}
\cite{amoros1} based on the data
\footnote {An indicative fit using preliminary data from
the more recent E865 experiment \cite{newkl4} is performed in ref. 
\cite{amoros1} which gives essentially the same central values for
$L_1$, $L_2$, $L_3$ and error bars reduced by approximately a factor
of two.}
of Rosselet et al. \cite{rosselet}. 
We note that the value of $L_1$ obtained previously in ref. 
\cite{Kl4rigg}, $10^3L_1=0.65\pm0.27$, while compatible, has a somewhat
larger central value.
Our results for $L_2$ and $L_3$ agree within approximately 30\% 
with the results from $K_{l4}$ or $\pi\pi$. 
Concerning $L_2$, 
the difference beween the $K_{l4}$ and the  $\pi\pi$ result is substantially
larger than that, 
our result happens to lie in between these two. 
A discrepancy at the 30\%
level can be expected as a consequence of unaccounted for $O(p^6)$ effects. 
In the case of $L_1$, however, there is a larger discrepancy, by about
a factor of two,  between our value and that from ref.\cite{amoros1}.
The combination $L_2-2L_1$ is suppressed in the large $N_c$ limit\cite{gl85}. 
We indeed find a  suppression of the value of $L_2-2L_1$ (compared, say, with 
$L_1(m_\rho)$ or  $L_2(m_\rho)$) even though this combination is dominated
by scalar resonances. A calculation of the $K_{l4}$ from factors in ChPT
to order $p^6$ was recently performed \cite{amoros1,amoros} and the couplings
$L_1,\ L_2,\ L_3$ were then redetermined, using a model to estimate the
$O(p^6)$ couplings $C_i(\mu)$. The following numbers are obtained 
\cite{amoros1}
\be
10^3 L_1= 0.53\pm0.25,  \quad
10^3 L_2= 0.71\pm0.27,  \quad
10^3 L_3=-2.72\pm1.12  \quad[K_{l4},\ O(p^6)]\ .
\en
Clearly, variations larger than
naively expected can occur as compared to the $O(p^4)$ determination. It 
remains to be seen, and this would be an interesting check of the convergence
of the $SU(3)$ chiral expansion, how the differences in the results from
the various methods of determining  $L_1,\ L_2,\ L_3$ are reduced, 
once the $O(p^6)$ contributions are included. 

In order to estimate the errors, firstly, we have varied the $S$ and $P$
wave $\pi K$ phase-shifts inside bands of half-width 
$\Delta\delta_0^{1/2}=2^\circ$, $\Delta\delta_0^{3/2}=1^\circ$, 
$\Delta\delta_1^{1/2}=1^\circ$, which correspond to the average experimental
errors in the region of elastic scattering (which makes the 
most important contribution). 
For the $\pi\pi\to K\overline K$ P-wave, $g_1^1$, we have varied the
coefficients of the normalising polynomial (see eq. \rf{g11pol}), allowing
for a term linear in $t$, and requiring that the $\chi^2$ does not exceed
twice its minimal value. The coefficient $\alpha_1$ is kept fixed since
its variation can be considered as an $O(p^6)$ effect, which we do not
try to estimate. This procedure generates a variation of the height of the
$\rho$ resonance peak in the 10\% range, which may seem rather small, 
but affects the results quite substantially, as can be seen from table 3.
For the $S$-wave, $g_0^0$, 
we have varied the phase at threshold $\phi_S(4m^2_K)$
(which, as we have seen, is the parameter on which the size of the 
$f_0(980)$ peak mostly depends) 
in a range between 150 and 200 degrees. Above threshold we have made 
$\vert g_0^0\vert$ to vary in the whole range allowed by the two incompatible 
experiments. We have also allowed a 10\% variation of 
the scattering lengths $a_0^{1/2}$, $a_0^{3/2}$ (keeping, however, the 
difference fixed) and, finally, in the Regge region we have assumed a 100\%
uncertainty. We show the individual impact of these variations on the
sum rule results in table 3. 
\begin{table}[hbt]
\begin{center}
\begin{tabular}{|c|c|c|c|l|}\hline
$10^5\times $   &$\Delta L_1$&$\Delta L_2$ &$\Delta L_3$ &$\Delta L_4$ 
\\ \hline\hline
$\delta^{3/2}_0$&0.2 & 0.5 & 0.7 & 0.5 \\ \hline
$\delta^{1/2}_0$&0.7 & 0.  & 3.0 & 3.6 \\ \hline
$\delta^{1/2}_1$&0.1 & 1.5 & 2.5 & 0.6 \\ \hline
$a_0^{1/2}$     &0.3 & 0.4 & 1.4 & 0.6 \\ \hline
$g_1^1$         &9.  & 9.  & 36. & 10. \\ \hline
$g_0^0$         &4.5 & 0.  & 0.  & 15. \\ \hline 
Regge           &0.3 & 1.6 & 1.0 & 0.1 \\ \hline
\end{tabular}
\end{center}
\caption{\sl  List of different sources of errors (see text for details)
and their impact on the determination of the $L_i$'s.}
\label{Table 3}
\end{table}

We now come to the discussion of $L_4$. From  relation \rf{L4} an important
remark can be made concerning the convergence: while $\beta^+$ and $\beta^-$
, separately, contain integrals which are slowly convergent at infinity, 
$L_4$ involves the difference, which has much better convergence properties.
Indeed, consider the high-energy contribution
\be\lbl{L4asy}
[\beta^+ -\beta^-]_{HE}= G_b^+(2)-G_b^-(2)-( H_b^+(2)+H_b^-(2))+
2\Skpi (H^+(3)-2 H_b^+(3)-H_b^-(3)).
\en
On rather general grounds, the leading Regge contribution in $G_b^+(2)$
and $G_b^-(2)$ is the same and will cancel out in the difference. Also in
the second potentially dangerous term $H_b^+(2)+H_b^-(2)$ the relevant
cross channel is pure $I=3/2$ and has no leading Regge contributions. The
other terms in eq. \rf{L4asy}  are, as we have seen, rather insensitive to the
asymptotic region. Therefore, we expect the uncertainty in $L_4$ 
coming from
the asymptotic region to be small, of the same size as in $L_1$, $L_2$,
$L_3$. Let us now consider the energy region below 1 GeV. Using 
eqs. \rf{L4} and \rf{sr} we can write,
\be\lbl{L4le}
[L_4]_{LE}=2A_1^++2\hat A_1^+-B_0^+ - B_0^- +v_0 -u_1,
+2\Skpi(-C_0^+ - C_0^- +w_0)
\en 
(where we have used $ A_1^- =A_1^+$, $\hat A_1^- =\hat A_1^+$ which is
true up to $O(p^8)$). This expression contains P-wave contributions,
which may seem surprising in view of the well known
resonance saturated expression \cite{egpr}
\be
L^{res}_4=-{c_d c_m\over3M^2_{S_8}}+{\tilde c_d \tilde c_m\over M^2_{S_1}}
\en
which involves only scalar resonances. It is in fact possible to write an
alternative expression for \rf{L4le}: using the crossing symmetry relation
\rf{cross} the P-wave combination $2\hat A_1^+ -u_1$ gets replaced by 
contributions from above the resonance region and the P-wave term $A_1^+$
has, in fact, no contribution from the resonances. This alternative
expression has only S-wave resonance contributions but is not as 
rapidly convergent.   

Numerical results for $L_4$ are shown in table 4 for several input values
of the threshold phase $\phi_S(4m^2_K)$ . 
We have
also mentioned that the two experimental measurements of 
$\vert g_0^0\vert$
of refs. \cite{cohen} and \cite{longacre} have a 
somewhat inconsistent normalisation.
We have performed the calculation for each data set separately.
\begin{table}[hbt]
\begin{center}
\begin{tabular}{|c|c|c|c|c|c|l|}\hline
$\phi_S(4m^2_K)$&$150^\circ$&$175^\circ$&$185^\circ$&$200^\circ$&
$220^\circ$&\ \\ \hline
$10^3\ L_4$&$ 0.08$&0.18&0.22&0.27&0.34&Etkin\\ 
           &$ 0.03$&0.10&0.13&0.16&0.21&Cohen      \\ \hline
\end{tabular}
\end{center}
\caption{\sl  Sum rule results for $L_4(\mu=0.770)$.}
\label{Table 4}
\end{table}
A clear feature from these calculations is that $L_4(m_\rho)$ is 
suppressed, and has a magnitude similar to $2L_1-L_2$ . It is not very easy
to decide on which central value to choose. We will make the choice 
of believing
the data of Cohen et al. \cite{cohen} for  the value of $\phi_S(4m^2_K)$
, which is then close to $200^\circ$, as they argue that the presence
of a P-wave in their experiment (which is absent in the other experiment)
helps in correctly determining the S-wave phase at threshold.
Concerning the normalisation of
$\vert g_0^0\vert$ above the threshold we may average over both experiments.
Taking into account   the
main sources of uncertainty (see table 3) we would then obtain,
\be\lbl{L4res}
L_4(\mu=0.770)=(0.22\pm 0.30 )\cdot 10^{-3}\ .
\en
This result is a refinement of the previous estimate of Gasser and 
Leutwyler \cite{gl85}, $L_4(\mu=0.770)=(-0.3\pm 0.5)\cdot 10^{-3}$, based on
the assumption that OZI suppression holds but without precisely knowing
the value of the scale at which it does. Other results can be found in
the literature \cite{moi,oller} which, however, are based on some
assumptions allowing one to determine scalar form-factors. 
The fact that they agree with \rf{L4res} indicates that 
these assumptions are reasonable. Finally, Amoros et al. \cite{amoros1} have
attempted to  determine $L_4$ from $K_{l4}$ data, using their $O(p^6)$
calculations, and they obtain $L_4(\mu=0.770)=(-0.2\pm 0.9)\cdot 10^{-3}$ 
which, as expected, is not very tightly constrained.

\noindent{\large\bf 5. Conclusions}

In this paper we have performed an evaluation of the set of sum rules
proposed in ref. \cite{ab}. While some results were already presented
in ref. \cite{ab}, the calculations performed here are more complete: they
take into account the contributions from partial waves beyond the S and P
waves as well as asymptotic contributions. In order to fully exploit these
sum rules, one needs to perform an extrapolation of the $l=0$ and $l=1$
partial waves of the $\pi\pi\to K\overline K$ amplitude. This can be performed
using standard Muskhelishvili-Omn\`es techniques and was considered
a long time ago \cite{J&P}. A great improvement over these calculations is
the availability nowadays 
of direct and precise experimental results concerning the 
$\pi\pi\to K\overline K$ amplitude, so there is no need to make use of
inelasticity in $\pi\pi$ scattering, which is not very precisely determined,
and requires the assumption of exact two-channel unitarity. We have checked 
the stability of the calculation by comparing  different
approaches to the solution. The main source of uncertainty comes from the
value of the phase at the $K \overline K$ threshold because 
the height of the $f_0(980)$
peak is strongly correlated with the value of this phase.
The two available experimental
data sets do not agree on this value and we have assumed a plausible range
of variation.
We have obtained a redetermination of the three $O(p^4)$ chiral
couplings $L_1$, $L_2$, $L_3$. Comparison with former determinations
allows a test of the $SU(3)$ chiral expansion. For instance, 
it is encouraging that the value
of $L_2$ that we obtain is intermediate between the value from $K_{l4}$
and the value from $\pi\pi$ sum rules. Besides, we have obtained for
the first time an evaluation of $L_4$ at the same level of precision
and reliability as $L_1$, $L_2$, $L_3$. This was not possible from
the $K_{l4}$ from-factors because of an accidental suppression of the
coefficient of $L_4$ in this case.

In constructing the $l=0$ and $l=1$
partial waves of the $\pi\pi\to K\overline K$ amplitude, we have solved
a subset of the system of Roy-Steiner equations. A further improvement,
which we have not performed here, would be to use the full set of equations
in order to constrain the low energy part of the $\pi K\to \pi K$ 
amplitudes. We note however, that the range of energies where this is needed,
that is, between the threshold and the energy where the data start is
smaller for $\pi K$ than it is for $\pi\pi$. Solving these equations
would  help in deciding whether a strange counterpart of the
$\sigma$ meson, the $\kappa$ meson, actually exists 
(e.g. ref. \cite{black} and references therein) or
not \cite{cherry}. An obvious further improvement would be to use the sum rules
in association
with a chiral $O(p^6)$ calculation of the $\pi K$ amplitudes. These, taken
together with the $\pi\pi$ sum rules (associated also with an $O(p^6)$ $SU(3)$ 
calculation of the $\pi\pi$ amplitude) 
and the available calculation
of the $K_{l4}$ form-factors to this order \cite{amoros} would no doubt
greatly improve our understanding of the chiral expansion in $m_s$.

\noindent{\bf Acknowledgements}
B.M. would like to thank B. Nicolescu for tuition on Regge physics,
M. Pennington for correspondence, W. Dunwoodie 
and R.S. Longacre for providing informations on the data.
P.B. would like to thank IPN Orsay for its hospitality and financial 
support during his stay in Paris.    

This work is supported in part by the EEC-TMR contract ERBFMRXCT98-0169,
by IFCPAR contract 2504-1 and DFG under contract np. ME 864-15/2.

\end{document}